\documentclass[intlimits,twoside,a4paper]{article}
\usepackage{graphicx}
\usepackage{amsmath,amssymb}
\usepackage[T2A]{fontenc}
\usepackage[cp1251]{inputenc}
\usepackage{cmpj}

\issue{2011}{14}{1}{13001}

\doinumber{10.5488/CMP.14.13001}

\title[Multidimensional thermodynamic potential for descriptions of ultrathin lubricant film melting]%
{Multidimensional thermodynamic potential for descriptions of ultrathin lubricant film melting \\ between two atomically smooth surfaces%
}
\author[L.S.~Metlov, A.V.~Khomenko, I.A.~Lyashenko]{L.S.~Metlov\refaddr{label1},
        A.V.~Khomenko\refaddr{label2}, I.A.~Lyashenko\refaddr{label2}}
\addresses{
\addr{label1} Donetsk Institute for Physics and Engineering named after O.O.~Galkin of the
National Academy of Sciences of Ukraine, 72 Rosa Luxemburg Str., 83114 Donetsk, Ukraine
\addr{label2} Sumy State University, 2 Rimskii-Korsakov Str., 40007 Sumy, Ukraine
}

\authorcopyright{L.S.~Metlov, A.V.~Khomenko, I.A.~Lyashenko}
 \date{Received January 19, 2010, in final form April 6, 2010}

\begin{document}

\maketitle

\begin{abstract}

The thermodynamic model of ultrathin lubricant film melting, confined between
two atomically-flat solid surfaces, is built using the Landau phase transition
approach. Non-equilibrium entropy is introduced describing the part of thermal
motion conditioned by non-equilibrium and non-homogeneous character of the
thermal distribution. The equilibrium entropy changes during the time of
transition of non-equilibrium entropy to the equilibrium subsystem. To describe the condition
of melting, the variable of the excess volume
(disorder parameter) is introduced which arises due to chaotization of a
solid structure in the course of melting. The thermodynamic and shear melting is
described consistently. The stick-slip mode of melting, which is observed in experiments, are described.
It is shown that with growth
of shear velocity, the frequency of stiction spikes in the irregular mode
increases at first, then it decreases, and the sliding mode comes further
characterized by the constant value of friction force. Comparison of the
obtained results with experimental data is carried out.
\pacs 05.70.Ce, 05.70.Ln, 47.15.gm, 62.20.Qp, 64.60.-i, 68.35.Af
\end{abstract}

\section{Introduction}\label{sec1}

With the development of nanotechnologies, the friction of two smooth
solid surfaces at the presence of thin lubricant film between them has been
vastly investigated lately~\cite{Persson}. Experimental
study of atomically-flat surfaces of mica, which are separated by
ultrathin layer of lubricant, shows that lubricant has
properties of solid at certain conditions~\cite{Yosh}. In
particular, the interrupted motion (\textit{stick-slip})  inherent in a dry friction~\cite{SRC,Aranson} is
observed. Such
boundary mode is realized, if the film of lubricating material has
less than four molecular layers, and this is explained as
solidification conditioned by  compression of walls. The
subsequent jump-like melting takes place, when shear stress
exceeds a critical value due to the effect of ``shear melting''.

In a general case, the description of behavior of ultrathin
lubricant film should be carried out starting with the first
principles. However, such an approach is complicated due to
different lubricants and geometry of experiment used. Therefore,
the phenomenological models are proposed that allow us to explain
the experimentally observed results. In particular,
thermodynamic~\cite{Popov},
mechanistic~\cite{Carlson,Filippov,Filippov1}, and
synergetic~\cite{KhYu} models are built. They are of
deterministic~\cite{Carlson,KhYu} and
stochastic~\cite{Filippov,Filippov1} nature. Also, the studies are
carried out using methods of molecular
dynamics~\cite{Braun,prd,Khome2010,SS_2010}. It appears that the
lubricant can provide several kinetic modes~\cite{Yosh}, between
which transitions occur stipulating the interrupted
friction~\cite{Yosh,Aranson}. In work~\cite{Filippov} three modes
of friction are revealed at the account of stochastic effects:
sliding mode that is inherent in low-velocity shear, regular
interrupted mode, and mode of sliding at high-velocity shear. The
existence of these modes is also confirmed by the synergetic
theory taking into account the deformation defect of the shear
modulus of lubricant~\cite{JPS} and the computer
experiments~\cite{Persson,Yosh,SRC,silica}.

In work~\cite{KhYu} within the framework of the Lorenz model for
approximation of viscoelastic medium, the approach is developed
according to which the transition of ultrathin lubricant film from
the solid-like state to the liquid-like state takes place as a
result of thermodynamic and shear melting. An analytical
description of these phenomena was presented under the assumption that
they are the results of shear stress and strain self-organization as
well as of the lubricant temperature. Additive noises of the
indicated quantities~\cite{dissipative,period,fnl_10} and
correlated fluctuations of temperature~\cite{uhl_fnl} are taken
into account. The reasons for the jump-like melting and hysteresis,
which are observed in experiments~\cite{exp1,exp_n,Isr_rev}, are
considered in works~\cite{JPS,FTT,PhysLettA}. Three stationary
modes are found out: two solid-like, inherent in the dry friction,
and liquid-like, corresponding to sliding. It is shown that
transition between two last modes takes place in accordance with
hysteresis of the dependence of stress on strain (at the jump-like
melting) or on the temperature of friction surfaces.

At the same time, the traditional use of the Lorenz equations set
for the above problem leads to logical contradictions already at
the stage of formulation of the problem. Consideration of strain
and stress as independent quantities, i.e., that for each of them
a separate kinetic equation is written, contradicts the canons of
``classic'' mechanics and thermodynamics. Moreover, there is no
symmetry of types of thermodynamic flows at such formulation which
predetermines strict accordance of signs in the mixed terms in
kinetic equations. An output can be found using a multidimensional
thermodynamic potential from which the set of Landau-like kinetic
equations must follow by standard procedure of differentiation
\cite{stat}. Earlier this approach was applied for description of
the processes of severe plastic deformation
(SPD)~\cite{m08,gm10,VDU4} and fracture of quasibrittle
solids~\cite{Metlov}. The latest advances and statistical
justification of this approach are outlined in
works~\cite{m1,m2,m3}. Both the process of SPD and the process of
ultrathin lubricant sliding have a lot in common, which allows us
to claim the legitimacy of application of such a technique in both
cases. However, principal differences between these processes are
present. They are mainly related to an ultrathin thickness of
lubricant layer (order of atomic size), which brings in the
limitations and deviations from standard procedure, which we shall
consider in the proper place of the article.

The general thermodynamic model of ultrathin lubricant film melting is built in the offered work.
Kinetic equations are written in as Landau-Khalatnikov ones for basic parameters (section~2).
In section~3 the effect of shear velocity is considered and it is shown that there exists a critical value
of velocity, at which lubricant melts in accordance with the mechanism
of shear melting. The effect of temperature is investigated. It is shown
that at temperature exceeding its critical value, the lubricant can melt even at the zero applied
shear stress and at zero velocity of the shearing, i.e., the thermodynamic melting
takes place (section~4). The effect of fluctuations of the strain is analyzed which
arise  due to errors in the experimental setup (section~5). All of the found
features coincide with the ones for experimental data.

\section{Thermodynamic model}\label{sec2}

At the construction of a model within the framework of the Landau
theory of phase transitions~\cite{stat}, at first, it is necessary
to choose a parameter whose values characterize the examined
phases. This parameter is called an order parameter and describes
a change of phase symmetry at the phase transition point. An order
parameter changes discontinuously during the first-order phase
transitions and it varies smoothly during the second-order phase
transitions. However, the phase symmetry changes discontinuously
at the phase transition point in both cases. A phase becomes
more ordered with the growth of the order parameter and symmetry falls
down.

Melting of a thin lubricant film unlike melting of volume medium can
take place according to the scenario of second-order phase
transition~\cite{Popov}. However, there is a certain problem at
describing the states of thin lubricant films, because such
films demonstrate more than one type of transition~\cite{Yosh}.
States of a lubricant film are not true thermodynamic phases.
They are interpreted as kinetic modes of friction~\cite{Aranson}.
Therefore, one speaks not about a solid and a liquid, but about a
solid-like and a liquid-like phases. The increase in
volume~\cite{Braun} and diffusion
coefficient~\cite{Braun,prd,thompson,liqtosol} of such lubricants
shows the melting process. Since the volume is experimentally observable,
to describe the state of lubricant, a parameter
$f$ is introduced, which also relates to the order parameter (a disorder
parameter~\cite{li}). It has a physical meaning of an excess volume,
arising due to the chaotization (the amorphization) of
structure of a solid during the process of melting. The excess
volume acquires zero value at zero Kelvin, when all atoms of the
system are densely packed at rest. It is different from zero both
in the solid and in the liquid state at non-zero temperature.
However, it has a larger value in the liquid state. We introduce
two values of this parameter: at $f>f_{\rm liq}$ lubricant is
liquid-like, and when $f<f_{\rm sol}$, it solidifies, and the
symmetry of state is decreased.

Now, in accordance with general procedure, it is necessary to write
down an expansion of energy in the independent variables. Internal
energy for a model, in which both contributions of large shear
strain $\varepsilon_{ij}^{\rm e}$ and of entropy $s$ and
nonequilibrium entropy $\tilde {s}$ are taken into account
simultaneously, is written as~\cite{VDU4}:
\begin{equation}
u = u_{0} + t_0 \tilde s - \frac{1}{2}t_1 \tilde {s}^2 + \varphi _0 f -
\frac{1}{2}\varphi _1 f^2 - m_1 \tilde s f,
\label{int_energy}
\end{equation}
where
\begin{eqnarray}
u_0 &=& \frac{1}{2}\lambda \left(\varepsilon_{ii}^{\rm e}\right)^2
+\mu\left( \varepsilon_{ij}^{\rm e}\right)^2 +
\alpha s^2 + \beta s\varepsilon_{ii}^{\rm e}\,,\\
\label{u_const}
t_0 &=& X_3^{(0)} \varepsilon_{ii}^{\rm e} + X_6^{(0)}
\left(\varepsilon _{ij}^{\rm e} \right)^2 + \alpha_t s,\\
\label{t_const}
\varphi_0 &=& g\varepsilon_{ii}^{\rm e} +
\bar\mu\left(\varepsilon _{ij}^{\rm e} \right)^2 + \alpha_\varphi s, \\
\label{varphi0_const}
\varphi_1 &=& \varphi_1^\ast + \tilde\mu\left(\varepsilon_{ij}^{\rm e}\right)^2, \\
\label{varphi1_const}
m_1 &=& m_1^\ast + X_3\varepsilon_{ii}^{\rm e} + X_6
\left(\varepsilon _{ij}^{\rm e}\right)^2 \label{m_const}
\end{eqnarray}
and $\lambda$, $\mu$, $\alpha$, $\beta$, $X_3^{(0)}$, $X_6^{(0)}$,
$\alpha_t$\,, $g$, $\bar\mu$\,, $\alpha_\varphi$\,,
$\varphi_1^\ast$\,, $\tilde\mu$\,, $m_1^\ast$\,, $X_3$\,, $X_6$
are constants of expansion.

Elastic stresses are taken into account with accuracy to quadratic
contributions via the first two invariants of the strain tensor
$\varepsilon_{ii}^{\rm e}$\,, $(\varepsilon_{ij}^{\rm
e})^2=\varepsilon_{ij}^{\rm e}\varepsilon_{ji}^{\rm e}$\,, where
summing is implied over repeated pairs of indices. Thus, the first
invariant represents the trace of the strain tensor
$\varepsilon_{ii}^{\rm e} = \varepsilon_1^{\rm
e}+\varepsilon_2^{\rm e}+\varepsilon_3^{\rm e}$\,, the second one
is determined by expression~\cite{Kachanov}
\begin{equation}
(\varepsilon_{ij}^{\rm e})^2 \equiv (\varepsilon_{ll}^{\rm e})^2 -
2I_2 = (\varepsilon_1^{\rm e})^2+(\varepsilon_2^{\rm
e})^2+(\varepsilon_3^{\rm e})^2.
\end{equation}
These determinations of invariants suppose that symmetric tensor
$\varepsilon_{ij}^{\rm e}$ of elastic strain is transformed to the
diagonal form.

A new basic quantity, i.e.,  the non-equilibrium entropy $\tilde {s}$, is
introduced here  describing the part of thermomotion, which is
conditioned by non-equilibrium character of the thermal
distribution. Exactly this part of the entropy is produced owing
to dynamic transition processes at generation of the free volume
during external action, tending to some stationary
value~\cite{m08,gm10,VDU4,Metlov,m1,m2}. The equilibrium entropy
does not evolve in the ordinary understanding, but changes with
time due to relaxation of  non-equilibrium entropy and its
transition into the equilibrium subsystem.

We write down the corresponding kinetic equations for
non-equilibrium parameters of state $X_i$ in the form
\begin{equation}
\tau_{X_i}\dot X_i = \frac{\partial u}{\partial X_i}\,,
\label{newton}
\end{equation}
where $\tau_{X_i}$ are the relaxation times.

At description by equations~(\ref{newton}), the system tends to a
maximum of the internal energy that corresponds to strongly
non-equilibrium processes occurring in open systems at energy
pumping therein (strictly speaking, this is true for an
``effective'' internal energy, which is a combination of the
internal energy and the entropic factor~\cite{m3}). For example,
the maximum of internal energy is of crucial importance for
magnetic~\cite{Vonsovski} and for alloy
orderings~\cite{klp91,gpdm03}. This property of the internal
energy is also similar to the property of thermodynamic potential,
introduced earlier for strongly non-equilibrium
processes~\cite{Panin}. In our case, the energy pumping is
realized due to deformation at the displacement of the friction
surfaces. Thus, a kinetic equation for the excess volume assumes
the form
\begin{equation}
\tau_f\frac{\partial f}{\partial t} = \varphi_0 - \varphi_1 f - m_1 \tilde s,
\label{kin_h}
\end{equation}
and for the non-equilibrium entropy $\tilde s$ we obtain
\begin{eqnarray}
\tau_s\frac{\partial \tilde s}{\partial t} = t_0 - t_1\tilde s - m_1 f,
\label{kin_s'}
\end{eqnarray}
where the terms with the sign~``+'' describe the increase in
non-equilibrium entropy due to the external source of energy (the
work), the terms with the sign~``--'' reflect its drift to
the equilibrium subsystem.

The kinetic equations for the equilibrium entropy differs from the
usual form~(\ref{newton}), since a change of equilibrium entropy
occurs due to transition of its non-equilibrium form to
equilibrium one.

In the case of non-homogeneous heating, the equation of heat
conductivity represents the ordinary equation of
continuity~\cite{Landau}:
\begin{equation}
T\frac{\partial s}{\partial t}=\kappa\nabla^2 T,
\label{teplo}
\end{equation}
where coefficient of heat conductivity $\kappa$ is a constant.
Supposing that a layer of lubricant and atomically-flat surfaces
have different temperatures $T$ and $T_{\rm e}$ accordingly, for a
normal constituent $\nabla^2_z$ it is possible to use the
approximation with sufficient accuracy $\kappa\nabla^2_z T \approx
(\kappa/h^2)(T_{\rm e} - T)$, where $h$ is the thickness of
lubricant. Taking this into account, equation~(\ref{teplo}) is
written down in a simple form
\begin{equation}
\frac{\partial s}{\partial t}=\frac{\kappa}{h^2}\left(\frac{T_{\rm
e}}{T}-1\right), \label{teplo_new}
\end{equation}
where quantity $h^2/\kappa$ plays the role of relaxation time, during which an
equalization of the temperature occurs over the thickness of the lubricant due to usual
heat conductivity.

The decrease in the non-equilibrium entropy is taken into account
by the negative terms in the kinetic equation~(\ref{kin_s'}), which
means that the same terms with positive sign must be taken into
account in the equilibrium entropy. Therefore, the final kinetic
equation for the equilibrium entropy is written down in the form:
\begin{equation}
\tau_s\frac{\partial s}{\partial t} = t_1\tilde s + m_1 f +
\frac{\tau_s\kappa}{h^2}\left(\frac{T_{\rm e}}{T}-1\right).
\label{kin_s}
\end{equation}
In accordance with the expression for the internal energy, the temperature of the lubricant is obtained in the form:
\begin{equation}
T = \frac{\partial u}{\partial s}  = 2\alpha s + \alpha_t\tilde s
+ \alpha_\varphi f + \beta\varepsilon_{ii}^{\rm e}\,. \label{T}
\end{equation}

According to~(\ref{int_energy}), elastic stresses are determined
as $\sigma_{ij}^{\rm e} = \partial u/\partial\varepsilon_{ij}^{\rm
e}$\,:
\begin{eqnarray}
\sigma_{ij}^{\rm e} &=& \lambda\varepsilon_{ii}^{\rm e}\delta_{ij}
+ 2\mu \varepsilon_{ij}^{\rm e} + \beta s \delta_{ij} + \left(
X_3^{(0)} \delta _{ij} +
2X_6^{(0)} \varepsilon_{ij}^{\rm e} \right)\tilde s  \nonumber \\
&+&\left(g\delta_{ij} + 2\bar\mu\varepsilon_{ij}^{\rm e} \right)f
- \tilde {\mu }\varepsilon _{ij}^{\rm e} f^2 -
\left(X_3\delta_{ij} + 2X_6\varepsilon_{ij}^{\rm e}\right)\tilde
{s}f. \label{sigma_ij}
\end{eqnarray}
Expression~(\ref{sigma_ij}) can be presented as the effective
Hooke law
\begin{equation}
\sigma_{ij}^{\rm e} = \sigma_{\rm v}\delta_{ij}+2\mu_{\rm
eff}\varepsilon_{ij}^{\rm e} + \lambda\varepsilon_{ii}^{\rm
e}\delta_{ij} \label{hooke}
\end{equation}
with the effective elastic parameter
\begin{equation}
\mu_{\rm eff} = \mu + X_6^{(0)}\tilde s + \bar\mu f -
\frac{1}{2}\tilde\mu f^2 - X_6\tilde s f\,. \label{mu_eff}
\end{equation}
Constants $\mu_{\rm eff}$ and $\lambda$ are the Lame
coefficients~\cite{Landau}. The term, being independent of strain,
appears in~(\ref{hooke})
\begin{equation}
\sigma_{\rm v} = \beta s + X_3^{(0)}\tilde s + gf - X_3\tilde s f.
\label{sigma_v}
\end{equation}
The first and second invariants are determined as
\begin{eqnarray}
\varepsilon_{ii}^{\rm e} &=& \frac{n-\sigma_{\rm
v}}{\lambda+\mu_{\rm eff}}\,,
\label{first_inv}\\
(\varepsilon_{ij}^{\rm e})^2 &\equiv& \varepsilon_{ij}^{\rm
e}\,\varepsilon_{ji}^{\rm e} =
\frac{1}{2}\left[\left(\frac{\tau}{\mu_{\rm eff}}\right)^2 +
\left(\varepsilon_{ii}^{\rm e}\right)^2\right], \label{second_inv}
\end{eqnarray}
where $n$, $\tau$ are the normal and tangential components of
stresses acting on the lubricant on the part of the rubbing
surfaces\footnote{Shear stress $\tau$ is defined from expression
(\ref{hooke}) at $i\ne j$, i.e., $\delta_{ij}=0$.}. The
relationships~(\ref{first_inv}) and~(\ref{second_inv}) represent
an ordinary connection between the components of tensors and their
invariants of linear elasticity theory (see, for
example,~\cite{Kachanov}). Let us use the Debye approximation
relating elastic strain $\varepsilon_{ij}^{\rm e}$ with plastic
one $\varepsilon_{ij}^{\rm pl}$~\cite{Popov}:
\begin{equation}
\dot\varepsilon_{ij}^{\rm pl} = \frac{\varepsilon_{ij}^{\rm
e}}{\tau_\varepsilon}\,. \label{e_pl}
\end{equation}
The total strain in a layer is determined as
\begin{equation}
\varepsilon_{ij} = \varepsilon_{ij}^{{\rm e}} +
\varepsilon_{ij}^{\rm pl}\,.
\end{equation}
This strain fixes the motion velocity of overhead block $V_{ij}$
according to relationship~\cite{Yosh}
\begin{equation}
V_{ij}=h\dot\varepsilon_{ij} = h (\dot\varepsilon_{ij}^{\rm e} +
\dot\varepsilon_{ij}^{\rm pl}). \label{V}
\end{equation}
Relaxation time of strain in~(\ref{e_pl}) depends on the state of
the lubricant:
\begin{equation}
\tau_\varepsilon = K(\gamma_0 - \gamma_1 f),
\label{tau_e}
\end{equation}
where the constants $\gamma_0$\,, $\gamma_1$ and coefficient $K$
are introduced. For the solid-like state of the lubricant $K =
K_{\rm sol}$\,.

In the solid-like state, $\tau_\varepsilon$ is large and, therefore,
$\varepsilon_{ij}^{\rm e}$ is large in accordance with the
expression~(\ref{tau_e}). For the liquid-like state,
$\tau_\varepsilon$ is small in comparison with the solid-like
case, and $\varepsilon_{ij}^{\rm e}$ is small too. Combining
relationships~(\ref{e_pl})--(\ref{tau_e}) we obtain an expression for
elastic shear strain:
\begin{equation}
\frac{\partial \varepsilon_{ij}^{\rm e}}{\partial t} =
-\frac{\varepsilon_{ij}^{\rm e}}{K(\gamma_0-\gamma_1 f)} +
\frac{V_{ij}}{h}\,. \label{e_ij_solid}
\end{equation}
The experimental data also evidence that in the liquid-like state,
the elastic strains relax rapidly~\cite{Yosh} and the
relaxation time for a liquid-like state is substantially
smaller. The expression~(\ref{tau_e}) at $K=K_{\rm sol}$ already
reflects the tendency of the relaxation time to decrease with
melting (at increase in $f$), but such a dependence is fulfilled
only for the solid-like state and near a transition
point~\cite{Popov}. Therefore, it is necessary to assume for the
liquid-like lubricant $K=K_{\rm liq}<K_{\rm sol}$\,.

It is known that the melting of lubricant is of hysteresis nature
in most cases~\cite{Filippov,exp1,exp_n,Isr_rev}. For theoretical
description of the hysteresis phenomena, a series of works were
undertaken, in particular, within the framework of Lorenz
model~\cite{JPS,FTT,PhysLettA}. In this approach, to account for
the indicated phenomena, it is necessary to select two
characteristic values of the excess volume: at $f>f_{\rm liq}$
lubricant is liquid-like, and when $f<f_{\rm sol}$ it solidifies.

Let us get an expression for the friction force that is measured
in experiments~\cite{Yosh}. Besides the elastic stresses
$\sigma_{ij}^{\rm e}$, the viscous ones $\sigma_{ij}^{\rm visc}$
also arise  in the lubricant. The total stress in a layer is the
sum of these two contributions
\begin{equation}
\sigma_{ij} = \sigma_{ij}^{\rm e} + \sigma_{ij}^{\rm visc}\,.
\label{sigma_all}
\end{equation}
The total friction force is determined in a standard manner:
\begin{equation}
F_{ij} = \sigma_{ij}A,
\label{F_begin}
\end{equation}
where $A$ is an area of contact. The viscous stresses in a layer
are given by the expression~\cite{Wear}
\begin{equation}
\sigma_{ij}^{\rm visc} = \frac{\eta_{\rm eff}V_{ij}}{h}\,,
\label{sigma_ij_v_begin}
\end{equation}
where $\eta_{\rm eff}$ is the effective viscosity that is defined
only experimentally, and for the boundary mode~\cite{Wear}
\begin{equation}
\eta_{\rm eff} \sim \left(\dot\varepsilon_{ij}\right)^\gamma.
\label{eta_eff}
\end{equation}
Taking into
account~(\ref{V}),~(\ref{eta_eff}), the expression for the viscous
stresses~(\ref{sigma_ij_v_begin}) is written down in the form:
\begin{equation}
\sigma_{ij}^{\rm visc} = \left(\frac{V_{ij}}{h}\right)^{\gamma+1}.
\label{sigma_ij_v}
\end{equation}
Putting~(\ref{sigma_all}),~(\ref{sigma_ij_v}) in~(\ref{F_begin}),
we obtain the final expression for the friction force:
\begin{equation}
F_{ij} = \left[\sigma_{ij}^{\rm e} +
\left(\frac{V_{ij}}{h}\right)^{\gamma+1}\right] A, \label{F}
\end{equation}
where $\sigma_{ij}^{\rm e}$ is fixed by the
expression~(\ref{hooke}) at $i\ne j$.

\section{The effect of velocity and shear melting}\label{sec3}

\begin{figure}[htb]
\centerline{\includegraphics[width=0.5\textwidth]{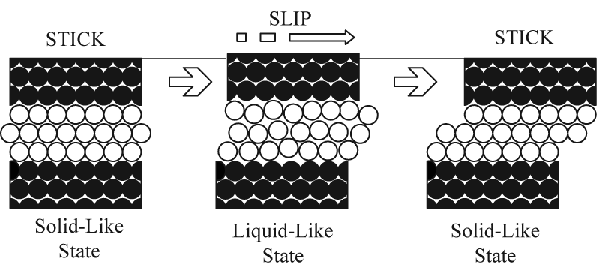}}
\caption{Schematic realization of the interrupted
(\textit{stick-slip}) friction mode~\cite{Yosh}.} \label{fig1}
\end{figure}

Ultrathin lubricant films behave differently from volume medium.
Therefore, at their description it is impossible to use standard
formalism without changes, since a number of principal new effects
appear, which must be taken into account. One of them is an
interrupted motion (\textit{stick-slip})~\cite{Yosh,Aranson}
schematically shown in figure~\ref{fig1}.
At the beginning, the lubricant is solid-like (\textit{stick}), but after
exceeding some critical value of the elastic shear stress
$\sigma_{ij}^{\rm e}$ it rapidly transforms into the liquid-like
phase (\textit{slip}) due to disordering. The top rubbing surface
climbs, because a change of lubricant volume takes place. The
relaxation of $\sigma_{ij}^{\rm e}$ occurs in the liquid-like
state and the lubricant solidifies again (\textit{stick}) due to
the compression by the walls under the action of a load. This
process is periodic. One of the basic differences from the
behavior of volume lubricants in this mechanism is that the action of
the shear stress $\sigma_{ij}^{\rm e}$ causes  not only a shear
but also an increase in lubricant volume. This fact agrees
with the results, which are obtained using the methods of molecular
dynamics~\cite{Braun}, and can be reflected by modifying the
expression~(\ref{first_inv}) as follows:
\begin{equation}
\varepsilon_{ii}^{\rm e} = \frac{n-\sigma_{\rm v}+\sigma_{ij}^{\rm
e}\,\varepsilon_{ij}^{\rm a}}{\lambda+\mu_{\rm eff}}\,.
\label{first_inv_new}
\end{equation}
The dimensionless tensor constant $\varepsilon_{ij}^{\rm a}$ is
introduced here, which fixes the dilatation power (the expansion
of lubricant at a shear under the action of $\sigma_{ij}^{\rm e}$).
Thus, it is also necessary  to take into account that the action of
shear stresses leads to an increase in lubricant thickness $h$.
The relative increase in volume\footnote{Physical meaning of the first
invariant~(\ref{first_inv_new}) is the relative change of volume
$\delta V/V_0$\,, where $\delta V$ is the change of volume, and
$V_0$ is the initial volume before deformation.} due to an increase
in lubricant thickness $h$ may be expressed by:
\begin{equation}
\frac{\delta V}{V_0} = \frac{A\delta h}{Ah} = \frac{\delta
h}{h}\,, \label{deltah_old}
\end{equation}
where $A$ is the area of contact. Equating a contribution to the
relative increase in volume from~(\ref{first_inv_new}) due to
shear stresses and the expression~(\ref{deltah_old}), we obtain
the change of lubricant thickness in the form
\begin{equation}
\delta h = h\frac{\sigma_{ij}^{\rm e}\,\varepsilon_{ij}^{\rm
a}}{\lambda+\mu_{\rm eff}}\,. \label{deltah}
\end{equation}
In subsequent calculations, the thickness $h$ in~(\ref{e_ij_solid})
it is necessary to replace by expression $h+\delta h$\,. Now the
model is complete, because together with the thermodynamic melting
we take  melting by shear into account.

\begin{figure}[htb]
\centerline{\includegraphics[width=0.5\textwidth]{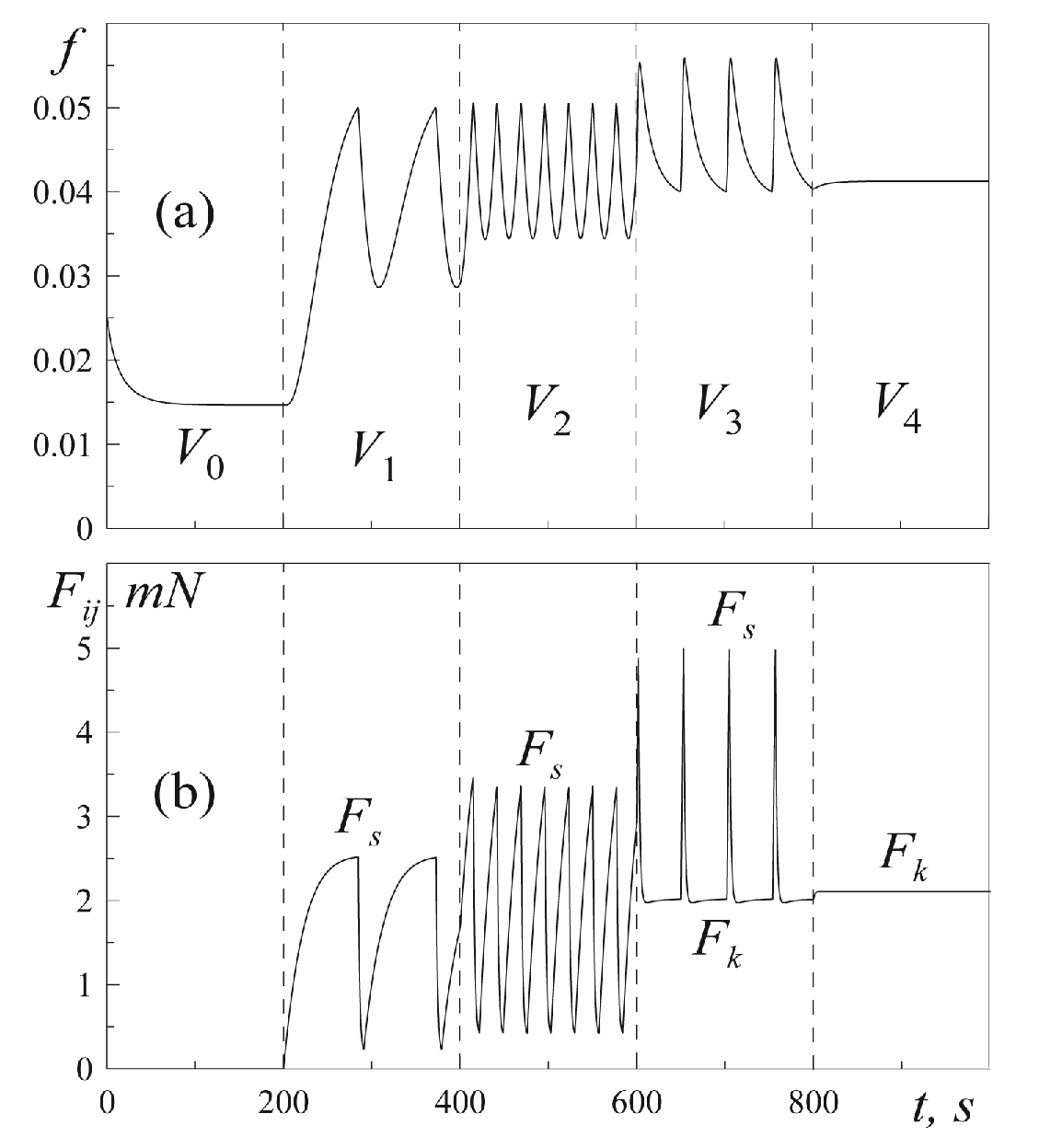}}
\caption{The time dependencies of excess volume $f$ and friction
force $F_{ij}$~(\ref{F}). Values of shear velocities $V_{ij}$ are
$V_0 = 0$~m/s, $V_1 = 1.9\cdot 10^{-12}$~m/s, $V_2 = 4\cdot
10^{-12}$~m/s, $V_3 = 21\cdot 10^{-12}$~m/s, $V_4 = 22\cdot
10^{-12}$~m/s.} \label{fig2}
\end{figure}

Result of simultaneous numerical solution of
equations~(\ref{kin_h}),~(\ref{kin_s'}),~(\ref{kin_s}),
and~(\ref{e_ij_solid}) is shown in figure~\ref{fig2} at parameters
$\lambda = 200~{\rm Pa}, \mu = 10^8~{\rm Pa}, \alpha = 4.1~{\rm
J^{-1}{\cdot}m^{3}{\cdot}K^2}, \beta = 2.2~{\rm K}, X_3^0 = 2\cdot
10^{-4}~{\rm K}, X_6^0 = 400~{\rm K}, g = 10^{-5}~{\rm Pa},
\bar\mu = 40~{\rm Pa}, \varphi_1^\ast = 5~{\rm J{\cdot}m^{-2}},
\tilde\mu = 2.4\cdot 10^{-3}~{\rm Pa}, m_1^\ast = 3\cdot
10^{-4}~{\rm K}, t_1 = 0.015~{\rm K}, X_3 = 1.12\cdot 10^{-4}~{\rm
K}, X_6 = 0.4~{\rm K}, \tau_f = 1~{\rm J{\cdot}m^{-3}{\cdot}s},
\tau_s = 0.1~{\rm J^{-1}{\cdot}m^3{\cdot}s{\cdot}K^2}, \kappa =
10^{-13}~{\rm Wt{\cdot}m^{-1}{\cdot}K^{-1}}, \alpha_t = 2\cdot
10^{-5}~{\rm J^{-1}{\cdot}m^{3}{\cdot}K^2}, \alpha_\varphi =
2\cdot 10^{-5}~{\rm K}, h = 10^{-8}~{\rm m}, T_{\rm e} = 300~{\rm
K}, \varepsilon_{ij}^{\rm a} = 1, n = -10^5~{\rm Pa}, \gamma_0 =
25~{\rm s}, \gamma_1 = 50~{\rm s}, K_{\rm sol} = 1, K_{\rm liq} =
0.07, f_{\rm sol} = 0.04, f_{\rm liq} = 0.05, A = 3\cdot
10^{-9}~{\rm m^2}, \gamma = 2/3$.
At zero velocity, the friction force is equal to zero, the excess volume $f$ decreases, lubricant here solidifies
slowly due to the compression by walls.

When the system begins motion ($V_{ij}=V_1\ne 0$), the lubricant
begins to melt under the action of growing stresses
$\sigma_{ij}^{\rm e}$, and the excess volume increases here. When
$f$ reaches the value $f=f_{\rm liq}$, the lubricant melts totaly,
and since the relaxation time in~(\ref{e_ij_solid}) becomes much
smaller, the stresses begin to relax. Here the lubricant begins to
solidify again, because the lubricant is supported by the elastic
stresses in the molten state. When it solidifies totally
($f=f_{\rm sol}$), due to the increase in the relaxation time
in~(\ref{e_ij_solid}), the parameter $f$ increases again, while it
does not reach the value $f_{\rm liq}$, and the process is reiterated
 again. According to this, the periodic interrupted
(\textit{stick-slip}) mode of melting/solidification sets in. It
should be noted that at $V_{ij}=V_1$ the excess volume at once
begins to decrease at exceeding the value $f_{\rm liq}$, while at
solidification and reaching $f=f_{\rm sol}$ it still decreases
during  some time, and only then it starts to increase. This is
because  some minimum value of stresses is needed for an increase
in $f$, and since the velocity is small, this value, according
to~(\ref{e_ij_solid}), slowly increases. Therefore, after
solidification, the excess volume can decrease during some time,
while the proper value of stresses is attained.

At an increase in velocity to the value $V_{ij}=V_2$, the frequency of stiction spikes
increases due to a rapid increase of stresses in the system at this velocity. Accordingly, the lubricant rapidly melts, and the system has got time to accomplish more transitions from
melting to solidification.

Frequency of peaks decreases again with the further increase in
velocity $V_{ij}=V_3$\,. This is because at high
velocity in equation~(\ref{e_ij_solid}), the stresses relax to a
greater stationary value at which the lubricant  solidifies slower.
The dependence $F_{ij}(t)$ has  long kinetic sections with
$F_{ij}=\rm const$. The excess volume increases during some time
in this mode at exceeding $f>f_{\rm liq}$ and then it begins to
decrease.

At further increase in shear velocity $V_{ij}=V_4$, the
interrupted mode disappears, and the kinetic mode of friction of
the liquid-like lubricant sets in with the value of friction force
$F_{\rm k}$\,. This takes place because at the value of velocity
more than critical one $V_{ij}>V_{\rm c}$ stress $\sigma_{ij}^{\rm
e}$ arises in a lubricant that provides a value $f>f_{\rm sol}$ at
which the lubricant is not capable of solidifying. Let us note
that at an increase in velocity, the values of stresses grow
according to the kinetic mode of friction with force $F_{\rm
k}$\,. This agrees with what is offered by a mechanistic
model~\cite{Carlson}.

Thus, at an increase in velocity, at first, frequency of stiction
spikes increases, then it decreases due to the appearance of long
kinetic sections, and when the critical value of velocity
$V_{ij}>V_{\rm c}$, is exceeded, the mode of \textit{stick-slip}
disappears. The described behavior well agrees with experimental
results~\cite{Yosh}.

\section{The effect of temperature}\label{sec4}

\begin{figure}[htb]
\centerline{\includegraphics[width=0.5\textwidth]{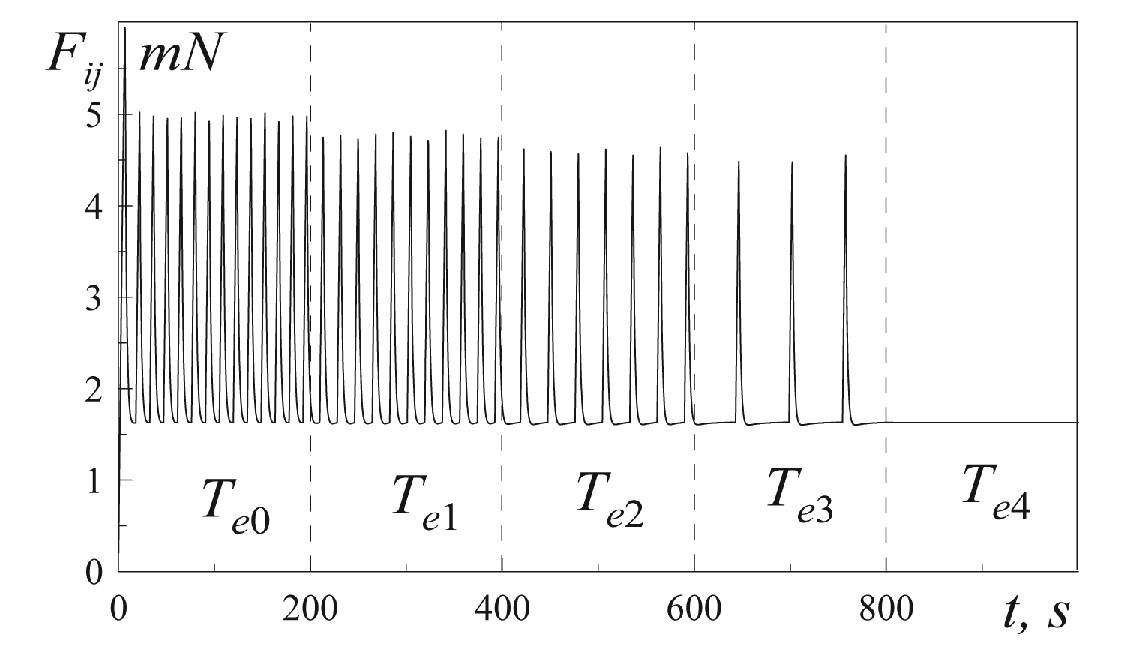}}
\caption{The time dependencies of the friction force $F_{ij}$
(\ref{F}) at parameters of figure~\ref{fig2} and shear velocity
$V_{ij}=17\cdot 10^{-12}$~m/s. Values of temperatures of friction
surfaces are $T_{{\rm e}0} = 50$~K, $T_{{\rm e}1} = 230$~K,
$T_{{\rm e}2} = 400$~K, $T_{{\rm e}3} = 490$~K, $T_{{\rm e}4} =
550$~K.} \label{fig3}
\end{figure}

The ultrathin lubricant films melt not only due to the shear
melting at the increase in velocity, but also due to the increase in
temperature. Let us investigate the effect of the
temperature on the examined system. For this purpose, we obtain
time dependencies for the friction force~(\ref{F}) which are
similar to the ones shown in figure~\ref{fig2}. Here, the value of shear
velocity $V_{ij}$ is assumed to be a constant, and the
temperature of the moved surfaces $T_{\rm e}$ increases. The
indicated dependencies are depicted in figure~\ref{fig3}.
It can be seen from the figure that at low temperatures of
friction surfaces $T_{\rm e} = T_{{\rm e}0}$, the frequency of
\textit{stick-slip} transitions is large, and the dependence
$F_{ij}(t)$ does not have the kinetic sections. This implies that
the lubricant begins to solidify immediately after melting. With
the increase in temperature ($T_{\rm e} = T_{{\rm e}1}$), both the
frequency of peaks and their height become smaller. Frequency
becomes smaller due to the appearance of kinetic sections, i.e.,
the lubricant now solidifies slower. A decrease in the height of
peaks implies the decline of static friction force $F_s$\,. With
the further increase in $T_{\rm e} = T_{{\rm e}2}$, the kinetic
section $F_{ij} = \rm const$ becomes more expressed, i.e., the
lubricant exists during some time in the molten state at constant
stresses which are already capable of supporting  this state.
However, due to dissipation the excess volume decreases and the
lubricant solidifies, and the \textit{stick-slip} mode is
realized. At $T_{\rm e} = T_{{\rm e}3}$, the kinetic section
becomes dominant, because most of the time the lubricant is in the
liquid-like state. At $T_{\rm e} = T_{{\rm e}4}$, the lubricant
melts ultimately and the kinetic mode is realized.

\section{The effect of noise}\label{sec6}

\begin{figure}[htb]
\centerline{\includegraphics[width=0.5\textwidth]{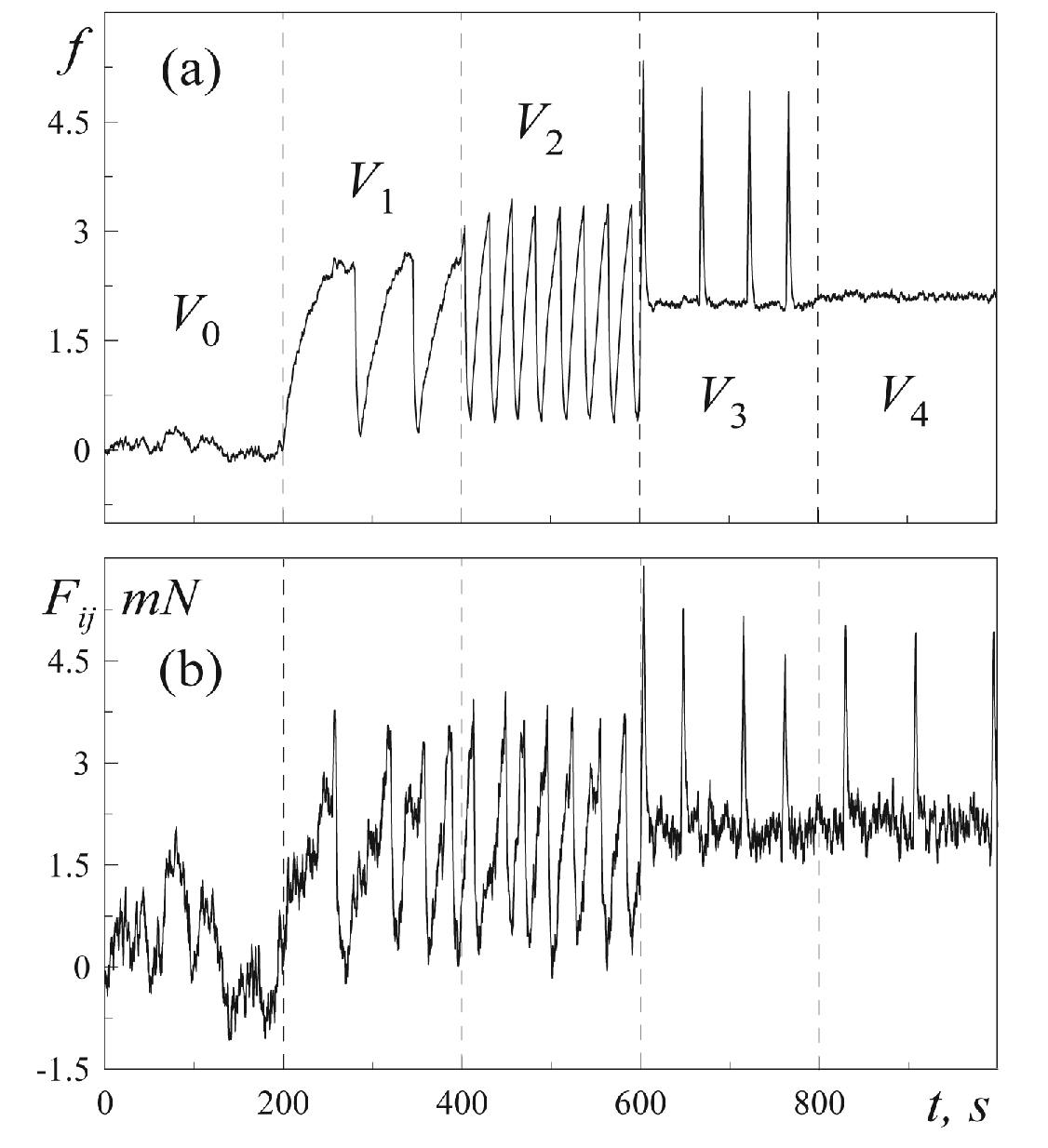}}
\caption{The time dependencies of friction force
$F_{ij}$~(\ref{F}) at parameters of figure~\ref{fig2} with
consideration of noise: a: $D=2\cdot 10^{-9}$~s$^{-1}$; b:
$D=8\cdot 10^{-8}$~s$^{-1}$.} \label{fig4}
\end{figure}

The deterministic case is considered above, though, in some
situations fluctuations critically effect the
system~\cite{dissipative,period,fnl_10}. We consider a case of
fluctuations appearing due to inaccuracy of experiment where the
value of the elastic strain is badly conserved
in~(\ref{e_ij_solid}), and thus it fluctuates. To this end, in the
right-hand part of~(\ref{e_ij_solid}) we add a stochastic source
$\xi(t)$ (the white noise mathematically defined using the Wiener
process~\cite{gard}), which has moments
\begin{equation}
\langle\xi(t)\rangle = 0,\qquad \langle\xi(t)\xi(t')\rangle =
2D\delta(t-t'), \label{xi_corr}
\end{equation}
where $D$ is the intensity of source. With this addition,
equation~(\ref{e_ij_solid}) has the form of Langevin equation:
\begin{equation}
\dot\varepsilon_{ij}^{\rm e} = -\frac{\varepsilon_{ij}^{\rm
e}}{K(\gamma_0-\gamma_1 f)} + \frac{V_{ij}}{h+\delta h} + \xi(t),
\label{langevin}
\end{equation}
whose solution within the framework of Ito prezentation with
account of~(\ref{xi_corr}) is carried out with the use of
iteration procedure in the form~\cite{dissipative,fnl_10}:
\begin{equation}
\varepsilon_{ij(n+1)}^e = \varepsilon_{ij(n)}^e +
\left(-\frac{\varepsilon_{ij}^e}{K(\gamma_0-\gamma_1 f)} +
\frac{V_{ij}}{h+\delta h}\right)\Delta t + \sqrt{2D\Delta t}W_n.
\label{iter}
\end{equation}
In order to model the random force $W_n$, the Box-Muller function is
used~\cite{dissipative,fnl_10,C++}:
\begin{equation}
W_n = \sqrt{-2\ln r_1}\cos(2\pi r_2), \quad r_i \in (0,1],
\end{equation}
where $r_1$ and $r_2$ are the pseudorandom numbers. The time
dependencies of the friction force~(\ref{F}) at the parameters of
figure~\ref{fig2} and using equation~(\ref{iter}) are shown in
figure~\ref{fig4}.
In figure~\ref{fig4}a, the intensity of noise is small and the
deterministic modes are realized~(cf. figure~\ref{fig2}b). With
the increase in intensity of fluctuations (figure~\ref{fig4}b), the
stochastic regime sets in at which the lubricant spontaneously
solidifies and melts. The possibility of existence of such modes was
shown by the methods of molecular dynamics~\cite{Braun}.

\section{Conclusions}\label{sec7}

The offered theory allows us to describe the effects observed at ultrathin
lubricant film melting. Both the ordinary case of  thermodynamic melting
due to an increase in temperature and the shear melting due
to disordering under the action of the applied external stresses are considered. It is shown
that these two processes are interconnected and they cannot be
examined separately. For example, at a high temperature of friction surfaces,
the shear melting is realized at smaller stresses, and at a still greater
increase in temperature, a lubricant melts even at zero stress (the
thermodynamic melting is realized).

The \textit{stick-slip} mode of friction, which is observed in
experiments, is naturally considered in the model, and it
is caused by the rapid relaxation of stress at reaching
the liquid-like state by a lubricant. At the temperature of friction surfaces
that does not provide the melting in a state of rest, the lubricant solidifies at such
relaxation again and it is solid-like during the
time that is necessary for the appearance of the stress at which melting
occurs. Thus, the temperature effect and shear melting are taken into
consideration. These are basic factors to be taken into account while
carrying out the experiments. The obtained dependencies
qualitatively coincide with the experimental ones.

\section*{Acknowledgements}

The work is partially executed with the financial support of the
Fundamental Researches State Fund of Ukraine (Grants
$\Phi$25/668--2007, $\Phi$25/97--2008, $\Phi$28/443--2009).

%
%

\ukrainianpart

\title{Багатовимірний термодинамічний потенціал для опису плавлення ультратонкої плівки мастила між двома атомарно-гладкими поверхнями}
\author{Л.С. Метлов\refaddr{label1}, О.В. Хоменко\refaddr{label2}, Я.О. Ляшенко\refaddr{label2}}
\addresses{
\addr{label1} Донецький фізико-технічний інститут ім.~О.О.~Галкіна НАН України, вул.~Р.~Люксембург,~72,  83114~Донецьк, Україна
\addr{label2} Сумський державний університет, вул.~Римського-Корсакова,~2, 40007~Суми, Україна
}
%
%
%

\makeukrtitle

\begin{abstract}
\tolerance=3000%
В межах теорії фазових переходів Ландау побудовано термодинамічну модель плав\-лення ультратонкої плівки мастила, що затиснута між двома
атомарно-гладкими твердими поверхнями. Введено нерівноважну ентропію, що описує частину теплового руху, який обумовлений
не\-рівноваж\-ним і нерівномірним характером теплового розподілу. Рівноважна ентропія змінюється у часі за рахунок переходу нерівноважної
ентропії в рівноважну підсистему. Для опису стану мастила введено параметр надлишкового об'єму (параметр безладу), який виникає
унаслідок хаотизації структури твердого тіла в процесі плавлення. Узгодженим чином описано термодинамічне і зсувне плавлення.
Описано переривчастий режим плавлення, який спостерігається в експериментах. Показано, що із зростанням
швидкості зсуву частота піків прилипання в переривчастому режимі спочатку збільшується, потім зменшується, і далі наступає режим
ковзання, що характеризується постійним значенням сили тертя. Проведено порівняння отриманих результатів з експерименталь\-ними даними.
\keywords межове тертя, нанотрибологія, динамічне моделювання, зсувні напруження та деформація, в'язкопружне середовище,
переривчастий режим
\end{abstract}


\begin{thebibliography}{99}
    \bibitem{Persson} Persson~B.N.J., Sliding Friction. Physical Principles and Applications.
        Springer-Verlag, Berlin, 1998.
    \bibitem{Yosh} Yoshizawa~H., Chen~Y.-L., Israelachvili~J.,
        J. Phys. Chem., 1993, \textbf{97}, 4128;
        \bibdoi{10.1021/j100118a033};
                   Yoshizawa~H., Israelachvili~J.,
        J.~Phys. Chem., 1993, \textbf{97}, 11300;
        \bibdoi{10.1021/j100145a031}.
    \bibitem{SRC} Smith~E.D., Robbins~M.O., Cieplak~M.,
        Phys. Rev. B, 1996, \textbf{54}, 8252;
        \bibdoi{10.1103/PhysRevB.54.8252}.
    \bibitem{Aranson} Aranson~I.S., Tsimring~L.S., Vinokur~V.M.,
        Phys. Rev. B, 2002, \textbf{65}, 125402;\\
        \bibdoi{10.1103/PhysRevB.65.125402}.
    \bibitem{Popov} Popov~V.L.,
        Tech. Phys., 2001, \textbf{46}, 605; \bibdoi{10.1134/1.1372955}.
    \bibitem{Carlson} Carlson~J.M., Batista~A.A.,
        Phys. Rev. E, 1996, \textbf{53}, 4153;
        \bibdoi{10.1103/PhysRevE.53.4153}.
    \bibitem{Filippov} Filippov~A.E., Klafter~J., Urbakh~M.,
        Phys. Rev. Lett., 2004, \textbf{92}, 135503;\\
        \bibdoi{10.1103/PhysRevLett.92.135503}.
    \bibitem{Filippov1} Tshiprut~Z., Filippov~A.E., Urbakh~M.,
        Phys. Rev. Lett., 2005, \textbf{95}, 016101;\\
        \bibdoi{10.1103/PhysRevLett.95.016101}.
    \bibitem{KhYu} Khomenko~A.V., Yushchenko~O.V.,
        Phys. Rev. E, 2003, \textbf{68}, 036110;
        \bibdoi{10.1103/PhysRevE.68.036110}.
    \bibitem{Braun} Braun~O.M., Naumovets~A.G.,
        Surf. Sci. Rep., 2006, \textbf{60}, 79;
        \bibdoi{10.1016/j.surfrep.2005.10.004}.
    \bibitem{prd} Khomenko~A.V., Prodanov~N.V.,
        Condens. Matter Phys., 2008, \textbf{11}, 615.
    \bibitem{Khome2010} Khomenko~A.V., Prodanov~N.V.,
        Carbon, 2010, \textbf{48}, 1234;
        \bibdoi{10.1016/j.carbon.2009.11.046}.
    \bibitem{SS_2010} Prodanov~N.V., Khomenko~A.V.,
        Surface Science, 2010, \textbf{604}, 730;
        \bibdoi{10.1016/j.susc.2010.01.024}.
    \bibitem{JPS} Khomenko~A.V., Lyashenko~I.A.,
        J. Phys. Stud., 2007, \textbf{11}, 268 (in Ukrainian).
    \bibitem{silica} Horn~R.G., Smith~D.T., Haller~W.,
        Chem. Phys. Lett., 1989, \textbf{162}, 404; \\
        \bibdoi{10.1016/0009-2614(89)87066-6}.
    \bibitem{dissipative} Khomenko~A.V., Lyashenko~I.A.,
        Tech. Phys., 2007, \textbf{52}, 1239;
        \bibdoi{10.1134/S1063784207090241}.
    \bibitem{period} Khomenko~A.V., Lyashenko~I.A.,
        Tech. Phys., 2010, \textbf{55}, 26;
        \bibdoi{10.1134/S1063784210010056}.
    \bibitem{fnl_10} Khomenko~A.V., Lyashenko~I.A., Borisyuk~V.N., FNL, 2010, \textbf{9},
    19;
    \bibdoi{10.1142/S0219477510000046}.
    \bibitem{uhl_fnl} Khomenko~A.V., Lyashenko~I.A.,
        FNL, 2007, \textbf{7}, L111;
        \bibdoi{10.1142/S0219477507003763}.
    \bibitem{exp1} Demirel~A.L., Granick~S.,
        J. Chem. Phys., 1998, \textbf{109}, 6889;
        \bibdoi{10.1063/1.477256}.
    \bibitem{exp_n} Reiter~G. et al.,
        J. Chem. Phys., 1994, \textbf{101}, 2606;
        \bibdoi{10.1063/1.467633}.
    \bibitem{Isr_rev} Israelachvili~J.,
        Surf. Sci. Rep., 1992, \textbf{14}, 109;
        \bibdoi{10.1016/0167-5729(92)90015-4}.
    \bibitem{FTT} Khomenko~A.V., Lyashenko~I.A.,
        Phys. Sol. State, 2007, \textbf{49}, 936;
        \bibdoi{10.1134/S1063783407050228}.
    \bibitem{PhysLettA} Khomenko~A.V., Lyashenko~I.A.,
        Phys. Lett. A, 2007, \textbf{366}, 165;
        \bibdoi{10.1016/j.physleta.2007.02.010}.
    \bibitem{stat} Landau~L.D., Lifshitz~E.M., Course of Theoretical Physics,
    Vol.5: Statistical Physics, 4th ed..
        Butterworth, London, 1999.
    \bibitem{m08} Metlov~L.S.,
         Bulletin of Russian Academy of Sciences, Physics 2008, \textbf{72},
         1283;\\  \bibdoi{10.3103/S1062873808090311}.
    \bibitem{gm10} Glezer~A.M., Metlov~L.S.,
        Phys. Sol. State, 2010, \textbf{52}, 1090;
        \bibdoi{10.1134/S1063783410060089}.
    \bibitem{VDU4} Metlov~L.S.,
        Bulletin of Donetsk Univ., 2007, \textbf{2}, 108 (in Russian).
    \bibitem{Metlov} Metlov~L.S.
        Preprint arXiv:cond-mat/0711.0399, 2007.
    \bibitem{m1} Metlov~L.S.
        Preprint arXiv:cond-mat.mess-hall/0910.5503, 2009.
    \bibitem{m2} Metlov~L.S.
        Preprint arXiv:physics.comp-ph/0912.2085, 2009.
    \bibitem{m3} Metlov~L.S.
        Preprint arXiv:cond-mat.stat-mech/1003.0450, 2010.
    \bibitem{thompson} Thompson~P.A., Grest~G.S., Robbins~M.O.,
        Phys. Rev. Lett., 1992, \textbf{68}, 3448;\\
        \bibdoi{10.1103/PhysRevLett.68.3448}.
    \bibitem{liqtosol} Gee~M.L., McGuiggan~P.M., Israelachvili~J.N.,
        J. Chem. Phys., 1990, \textbf{93}, 1895;
        \bibdoi{10.1063/1.459067}.
    \bibitem{li} Li~H.,
        Theory of phase transitions in disordered crystal solids.
        Ph.D. Thesis, Georgia Institute of Technology, 2009.
    \bibitem{Kachanov} Kachanov~L.M., Basics of Theory of Plasticity.
        Nauka, Moscow, 1969 (in Russian).
    \bibitem{Vonsovski} Vonsovskii~S.V.,
        Magnetism. John Wiley, New York, 1974.
    \bibitem{klp91} Kut'in~E.I., Lorman~V.L., Pavlov~S.V.,
        Sov. Phys. Uspekhi, 1991, \textbf{34}, 497; \\
        \bibdoi{10.1070/PU1991v034n06ABEH002385}.
    \bibitem{gpdm03} Gouyet~J-.F., Plapp~M., Dieterich~W., Maass P.,
        Adv. Phys., 2003, \textbf{52}, 523;\\
        \bibdoi{10.1080/00018730310001615932}.
    \bibitem{Panin} Panin~V.E., Yegorushkin~V.E., Khon~Yu.A., Elsukova~T.F.,
        Izvestiya VUZov, Physics, 1982, \textbf{12}, 5 (in
        Russian);  \bibdoi{10.1007/BF00900288}.
    \bibitem{Landau} Landau~L.D., Lifshitz~E.M., Course of Theoretical Physics, vol.7: Theory of Elasticity, 3rd ed..
        Pergamon Press, New York, 1986.
    \bibitem{Wear} Luengo~G., Israelachvili~J., Granick~S.,
        Wear, 1996, \textbf{200}, 328;
        \bibdoi{10.1016/S0043-1648(96)07248-1}.
    \bibitem{gard} Gardiner~C.W., Handbook of Stochastic Methods.
        Springer, Berlin, 1983.
    \bibitem{C++} Press~W.H. et al., Numerical Recipes in C: the Art of Scientific Computing.
        Cambridge University Press, New York, 1992.
\end{thebibliography}
\end{document}